\documentclass[journal]{IEEEtran}
\ifCLASSINFOpdf
\else
\fi
\usepackage[utf8]{inputenc}
\usepackage{booktabs}
\usepackage{multirow}
\usepackage{longtable}
\usepackage{graphicx}
\usepackage{setspace}
\usepackage{colortbl}
\usepackage[hidelinks]{hyperref}
\usepackage{svg}
\usepackage{amsmath}
\usepackage{mathtools}
\usepackage{caption}
\usepackage{graphicx}
\usepackage{float}
\usepackage{afterpage}
\usepackage{diagbox}
\usepackage{svg}

\usepackage{float}
\usepackage{subcaption}

\begin{document}
%
\title{The Quantum Tortoise and the Classical Hare:\\ 
A simple framework for understanding which problems quantum computing will accelerate (and which it won't)}
%
%
%

\author{Sukwoong Choi,
        William S. Moses,
        Neil Thompson
\thanks{Corresponding author: Sukwoong Choi}
\thanks{Sukwoong Choi is an Assistant Professor at the Department of Information Systems and Business Analytics, School of Business, in the University at Albany, State University of New York and a Digital Fellow at Initiative on the Digital Economy of MIT. Email: schoi27@albany.edu}
\thanks{William S. Moses is an Assistant Professor in the Computer Science department at the University of Illinois, Urbana-Champaign.\\ 
Email: \href{mailto:wsmoses@illinois.edu}{wsmoses@illinois.edu}}
\thanks{Neil Thompson is the Director of the FutureTech research project at MIT’s Computer Science and Artificial Intelligence Lab and a Principal Investigator at MIT’s Initiative on the Digital Economy. Email: neil\_t@mit.edu}}

\maketitle

\begin{abstract}
  Quantum computing promises transformational gains for solving some problems, but little to none for others. 
  For anyone hoping to use quantum computers now or in the future, it is important to know which problems will benefit. In this paper, we introduce a framework for answering this question both intuitively and quantitatively. The underlying structure of the framework is a race between quantum and classical computers, where their relative strengths determine when each wins. While classical computers operate faster, quantum computers can sometimes run more efficient algorithms. Whether the speed advantage or the algorithmic advantage dominates determines whether a problem will benefit from quantum computing or not. Our analysis reveals that many problems, particularly those of small to moderate size that can be important for typical businesses, will not benefit from quantum computing.  Conversely, larger problems or those with particularly big algorithmic gains will benefit from near-term quantum computing. Since very large algorithmic gains are rare in practice and theorized to be rare even in principle, our analysis suggests that the benefits from quantum computing will flow either to users of these rare cases, or practitioners processing very large data.
\end{abstract}


%
\IEEEpeerreviewmaketitle

\section{Introduction}
In the children's story of the Tortoise and the Hare, the speedier Hare is outpaced by a Tortoise with other advantages (diligence).  An analogous contest is happening in computing, between a Quantum Tortoise and a Classical Hare.  Here, the Classical Hare's speed advantage is literal --- classical computers run faster than quantum ones.  Like his namesake, the Quantum Tortoise is slower, but also has an advantage --- in this case, the ability to run algorithms that are unavailable to classical computers.  When this algorithmic advantage is important enough, the Quantum Tortoise can beat the Classical Hare and solve a problem faster.  This paper analyzes when the Quantum Tortoise will beat the Classical Hare --- and when it won't.

Understanding where quantum computing will provide improvements is important because quantum computing is often heralded as the next generation of computing --- the promising successor to traditional, classical computing that progresses ever more slowly as Moore's Law fades \cite{Preskill2012}. But the reality is more complicated.  In general, quantum computers are \textit{slower} than traditional ones and thus, all else equal, they are less capable. But all else is not equal. Quantum computers can run some algorithms that classical computers cannot. This is because quantum computers can represent large amounts of information in dense ``superpositions'', a feat that classical computers cannot achieve. This representation allows quantum computers to perform operations on the patterns between individual bits (superpositions), not just on the individual bits themselves.  When the algorithms unlocked by this difference provide enough of an improvement, quantum computers' speed disadvantage can be more than overcome.  

\begin{figure}[!ht]
    \centering
    \includegraphics[width=0.47\textwidth, scale=1]{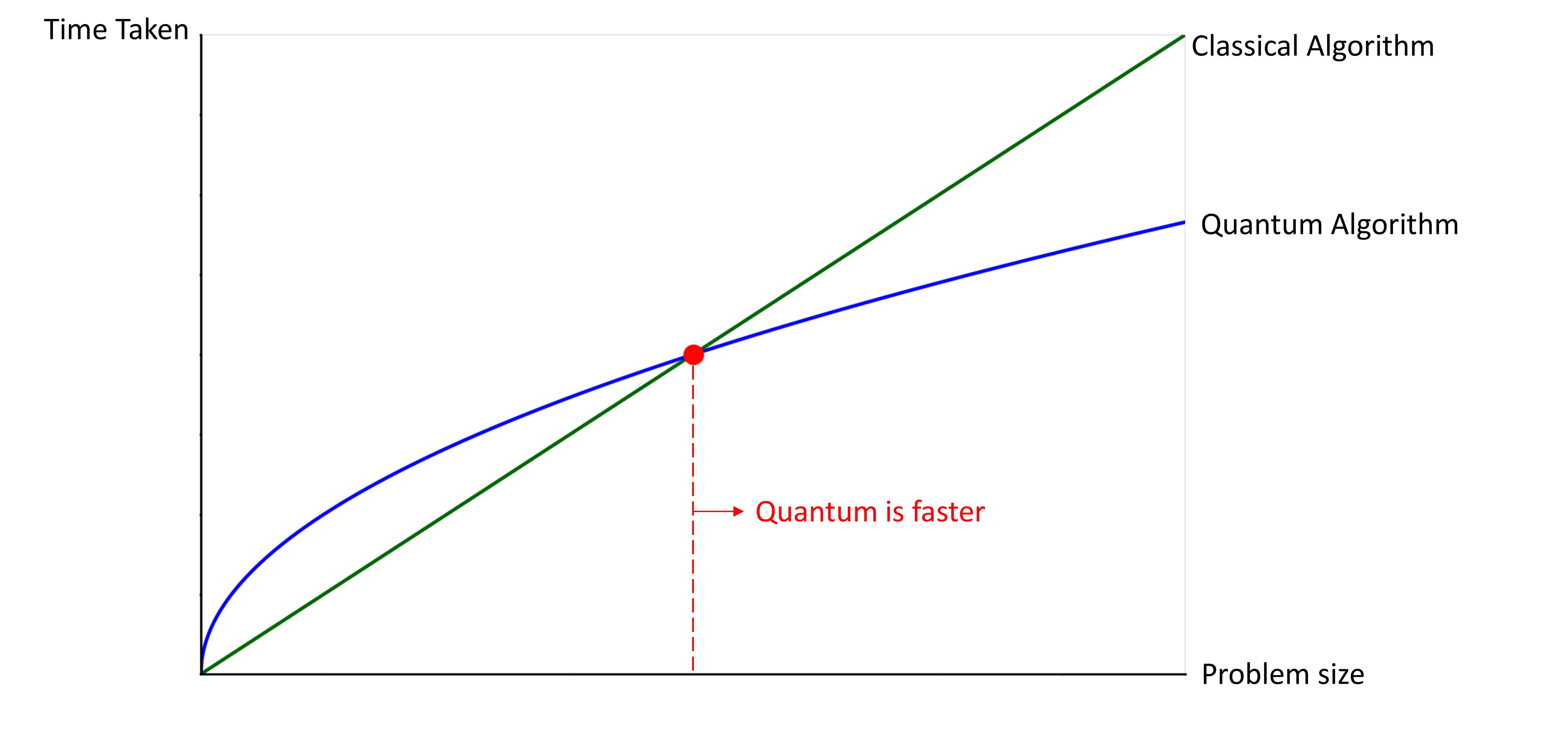}
        \caption{Conceptual representation of why a better quantum algorithm only outpaces a classical one when problem sizes are sufficiently large. Note: because quantum computers run more slowly, they can do many fewer calculations in one unit of time on the y-axis than can a classical computer.}
    \label{fig:conceptual}
\end{figure}

We find that there are two key determinants of whether a problem will be sped up: how much better the quantum algorithm is, and how big a problem is being solved (see \autoref{fig:conceptual}). The problem size matters because the benefit of an algorithmic advantage is larger for larger problems. This means that if a problem is too small, the classical computer will have already completed the problem by the time the quantum computer's algorithmic benefit kicks in. This means that problems with the potential for a quantum benefit will have a threshold point, where problem sizes below the threshold do not benefit from quantum, while problem sizes above do benefit.

Based on real estimates of where this threshold is for various types of problems, we find that quantum will be more attractive than classical computing in two principal cases, when (i) the classical algorithm takes an exponential amount of time and the quantum algorithm is faster, or (ii) the quantum algorithm is significantly better, say by a factor of $n$, and the problem size being tackled is larger than speed difference between classical and quantum computers --- for example if classical computers are $1{,}000{,}000\times$ faster then quantum computers will be better if the problem being tackled has $n > 1{,}000{,}000$.

\section{A framework for understanding where quantum computing will provide an advantage}

\subsection{Intuition}
Most predictions about the benefits of quantum are anecdotal extrapolations, for example referencing Shor's algorithm for factoring numbers (and thus breaking traditional cryptography ~\cite{rivest1978method}).  These predictions correctly point out the theoretical ability of quantum computers to tackle problems that are impossible on classical computers. For instance, it has been pointed out that with the fastest classical factorization algorithm, it would take longer than the age of the universe to factor 2048-digit RSA encryption, whereas Shor's quantum algorithm could theoretically solve it in days~\cite{jones2012layered}. There are two problems with such claims.  The first is timing. Current quantum computers are not powerful enough to do this calculation, and will not be for a while (the quantum record as of 2019 is factoring a 40-bit number, and that result also required substantial classical computation ~\cite{karamlou2021analyzing}). The second problem with such claims is that they only focus on tasks that are currently infeasible to compute. But it is also important to know whether quantum will speed up the tasks currently being done on classical computers. That is, will quantum just add new functionality or will it also improve computing more broadly? In this article, we construct a framework that addresses these concerns, focusing on near-term quantum computers and the problems they will be better equipped to solve, including those currently tackled by classical computers. We explain the framework in steps, starting with a discussion of cryptography and Shor's algorithm that is grounded in what computers are doing today.


Consider encryption that uses 2048-bits, as is common for modern computer systems~\cite{google_2013, sslpulse}. We can compare how long it takes to solve these problems and how this grows at the problem size (n) gets larger, as described by the time complexity of the algorithm used\footnote{For those not familiar with algorithmic notation, these express the scaling of algorithms: $O(n)$ means (roughly) that computational costs for classical computers increase linearly as problem size (n) grows, whereas $O(\sqrt{n})$ means that computational costs grow as the square-root of problem size.} to factor an $n$-bit number is $e^{(64/9)^{1/3}n^{1/3}(\log n)^{2/3}(1+o(1))}$, implying that the cost of factoring grows exponentially and that factoring a 2048-bit number would require roughly $10^{41}$ operations\footnote{Note: here we are assuming that the computational model used to derive the classical algorithm provides a faithful count for the number of operations on a real machine, and also assume a value of $1$ for the complexity overhead, and 0 for the inner $o(1)$ constant.}.
Extrapolating from the 2020 factoring result in \cite{nodacfactor}, one CPU-year can calculate $10^{24}$-$10^{25}$ factoring operations, meaning that factoring a 2048-bit number classically would take approximately $10^{16}$ cpu-years (or a million computers running for age of the universe~\cite{chaboyer1998age}).
So this problem is infeasible on a classical computer. What about on a quantum computer?

Using Shor's algorithm~\cite{shor1999polynomial}, a quantum computer can factor an $n$-bit number with time complexity $O\left(n^2 \log n \log \log n\right)$~\cite{beckman1996efficient}. This implies that a 2048-bit number would require roughly $10^7-10^8$ qubit operations. In reality, it would take more than this because of fault tolerance, error correction, and other overheads, but we will revisit corrections for these terms later.
Based on this optimistic estimate of qubit operations, and the current speed of quantum computers, approximately 2MHz\footnote{A simplified assumption based on the CNOT gate switching speed of IBM's eagle computer}, factoring a 2048-bit number would take less than a day\footnote{Under more realistic assumptions, one particular quantum architecture requires $10^9$ toffoli gates (approximately 10 days for 2012 state-of-the-art) and $10^5$ logical qubits to factor a 2048-bit number~\cite{jones2012layered}.}. So the running time on a quantum computer would be both reasonable and a great improvement on the classical computer. 
 But are the hardware requirements also reasonable?  Current quantum algorithms for factoring $n$-bit numbers require $O(n)$ qubits~\cite{anschuetz2019variational}, so such a computer would need about $10^3$ logical qubits. In practice, current quantum computers have very few logical qubits \cite{Ball2021}, so they cannot solve this problem. Indeed, even with generous extrapolations from current quantum roadmaps (discussed later), quantum computers would not be able to solve this problem until 2030. With less generous assumptions, it could be significantly later.

Hence, our framework suggests that quantum computers will be useful for the 2048-bit encryption problem, but only starting at the earliest in 2030.  What about for smaller encryption problems? The largest number publicly-known to have been factored happened in 2020, when a team in France and the US factored a 829-bit number in approximately 2700 CPU core-years using the number field sieve algorithm~\cite{nodacfactor,boudot2020comparing}. 
Revisiting our optimistic analysis, we find that a quantum computer can solve this 829-bit problem in under a day using significantly fewer logical qubits. So the smaller the number of bits used in the encryption, the earlier it will be feasible to construct quantum hardware to solve the problem faster. This is a more-general pattern: quantum will become useful sooner for problem sizes that offer a smaller benefit and only later become useful for problem sizes with larger benefits. This suggests a continuum, where even smaller problems don't get enough algorithmic advantage from quantum to be better than classical, but then there is a cross-over point where the problem is large enough for quantum to be better. Later, we will derive this explicitly.

\subsection{Framework}

To generalize the insights articulated in the example above we propose a framework that captures when you would want to preferentially use a quantum computer to solve a problem and actually could. This framework arises from two key constraints that are needed for quantum computers to be better. In what follows, we discuss these two constraints and how they lead to ``Quantum economic advantage,'' which we coin in a parallel construction to ``Quantum advantage.'' We develop this new terminology to reflect that traditional quantum advantage - which measures when a quantum computer can outperform any classical computer - will not be the right criterion for deciding whether to use them as the technology matures. In particular, as quantum computers improve there will be cases when they are solving problems slower than an expensive supercomputer, but faster than a comparably expensive classical computer. In such situations, the economically rational thing to do would be to use a quantum computer, even though it would \textit{not} have quantum advantage. Put another way, quantum advantage will underestimate the usefulness of quantum computers as they get cheaper than the most expensive classical computers. To resolve this problem, we add a modification to the definition of quantum advantage to express this difference:

\textbf{Definition: Quantum Advantage}\footnote{Adapted from Wikipedia "Quantum Supremacy" article.}
\begin{quote}
    Quantum computers have a \textit{\textbf{Quantum Advantage}} for a particular problem when a quantum computer exists that is capable of answering the question faster than \textit{any} classical computer.
\end{quote}    

\textbf{Definition: Quantum Economic Advantage}\footnote{Note: we use the term ``Quantum Economic Advantage'' differently than \cite{Bova2022}.}
\begin{quote}
    Quantum computers have a \textit{\textbf{Quantum Economic Advantage}} for a particular problem when a quantum computer exists that is capable of answering the question faster than \textit{a comparably expensive} classical computer.\footnote{More formally, there is a quantum economic advantage if there exists a quantum computer that can solve a problem instance for \$X and all existing classical systems with a computation budget \$X or less would solve the problem in more time.}
\end{quote}    

This quantum economic advantage definition incorporates two important conditions. First, a quantum computer must be sufficiently powerful to run the problem. 
 Second, the quantum computer must run it faster than a classical computer. That is, a problem must be \textit{feasible} and have sufficient \textit{algorithmic advantage}\footnote{We implicitly assume here that quantum computers are more expensive per cycle than classical computers. If this were not the case, a quantum computer would have a quantum economic advantage for all feasible problems.}.

\textbf{Feasibility.} 
A problem is \textit{\textbf{feasible}} for a quantum computer if that computer is sufficiently powerful to actually run the problem.

In today's hardware, the most important contributor to quantum feasibility is whether the system has sufficient qubits to run the problem using an implementable error correction. More problems will become feasible as qubit counts rise and error correction improves. As this definition makes clear, quantum feasibility is a constraint on which problems can be solved. This interpretation is shown schematically in Figure \ref{fig:QuantumEconomicAdvantage}(a).

\afterpage{
\begin{figure}[H]
    \centering
    \includegraphics[width=0.55\textwidth, scale=1]{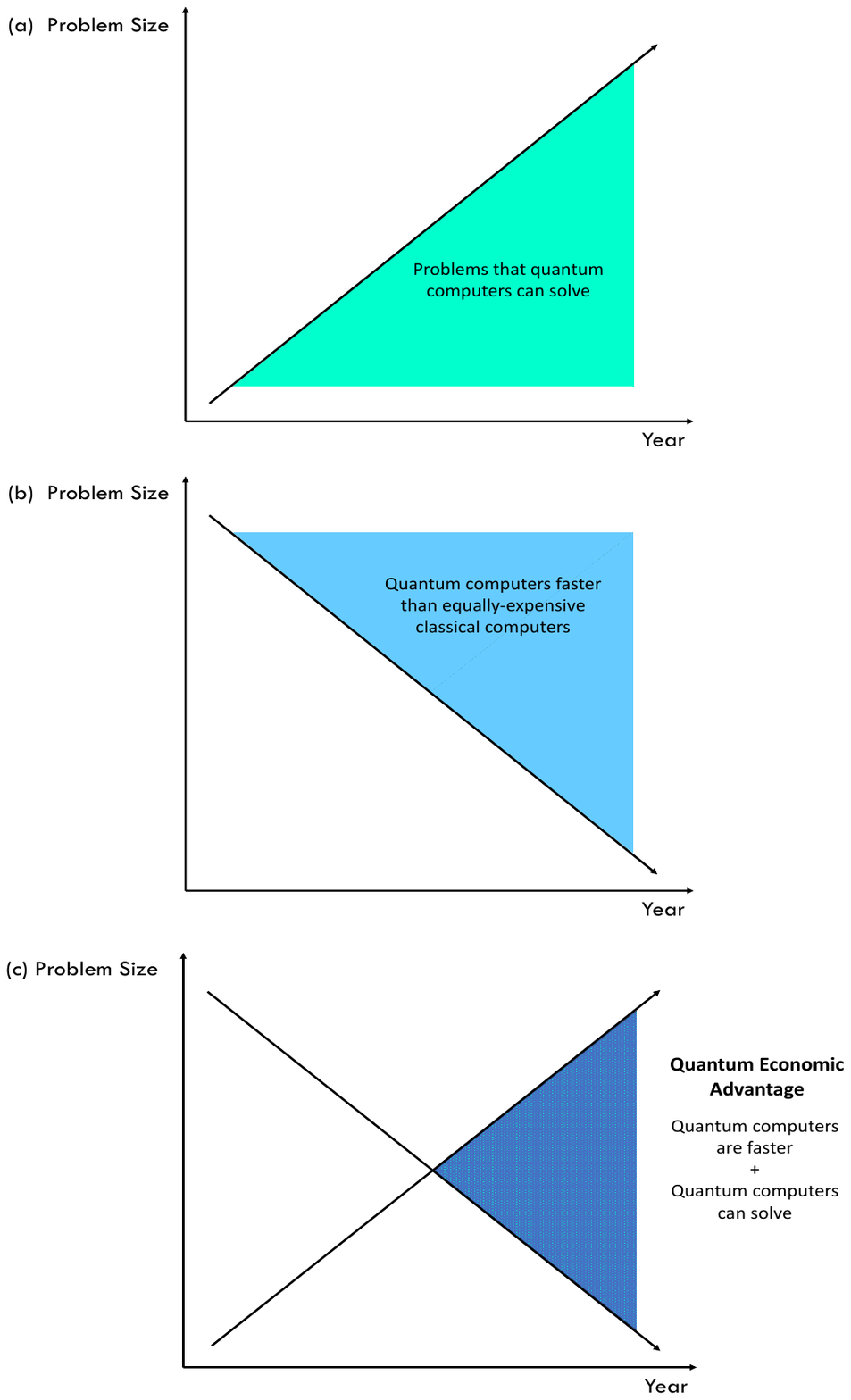}
    \caption{Schematic representation of when quantum economic advantage will occur, including (a) \textbf{Feasibility:} quantum computers are powerful enough to solve a particular problem size, and (b) \textbf{Algorithmic Advantage:} when a better quantum algorithm provides enough advantage to overcome the speed advantage of classical computers.  Collectively, these constraints produce (c) an overlap region of \textbf{Quantum Economic Advantage}.}
    \label{fig:QuantumEconomicAdvantage}
\end{figure}
}
\textbf{Algorithmic Advantage.}  A problem has a \textit{\textbf{algorithmic advantage}} if, for that size of problem, a quantum computer is faster at calculating it than a comparably expensive classical computer.

This race between a quantum computer and a comparably expensive classical computer is the one referred to in the article's title. A quantum computer has an algorithmic advantage when its superior algorithm provides enough advantage to overcome the speed advantage of the classical computer. It must also overcome any error correction overhead needed. To overcome these factors, a problem size will typically need to be sufficiently large. This is shown in the blue regions of Figure \ref{fig:QuantumEconomicAdvantage}(b). Here, we show the curve as being downward sloping, which assumes that quantum computers are getting better faster than classical computers over time.\footnote{This seems plausible since classical computer progress has slowed as Moore's Law ends \cite{leiserson2020there}, whereas quantum computers are early in development and improving rapidly (as we show in the hardware section). If this assumption were incorrect, and thus classical computers were to get better faster, then figure 1(b) would be upward sloping and only a very few, typically very large, problems would be better on a quantum computer.}


Combining these two constraints, as shown in the overlap region of Figure \ref{fig:QuantumEconomicAdvantage}(c), shows that quantum advantage will come in wedge-shaped patterns. More specifically, this framework suggests that problems will have some date when quantum computers will first be better.  Initially, this will only be true for some particular problem size that is small enough to fit on quantum computers of that year, but large enough to get sufficient advantage from the better algorithm.  After that date, the range of problem sizes ("quantum-advantaged problem sizes - QAPS") will grow as quantum computers get faster, cheaper and more capable.

\subsection{An example}

Consider the time needed to search a string of text for a particular sub-string, for example looking for a particular sequence of DNA in a large genome.  The best classical algorithm for this problem is $O(N)$ and the best quantum algorithm is $O(\sqrt{N})$ (Grover's algorithm).  This difference represents the algorithmic shortcut that our quantum tortoise gets.


\begin{figure*}[h]
    \centering
    \includegraphics[width=0.70\textwidth, scale=1]{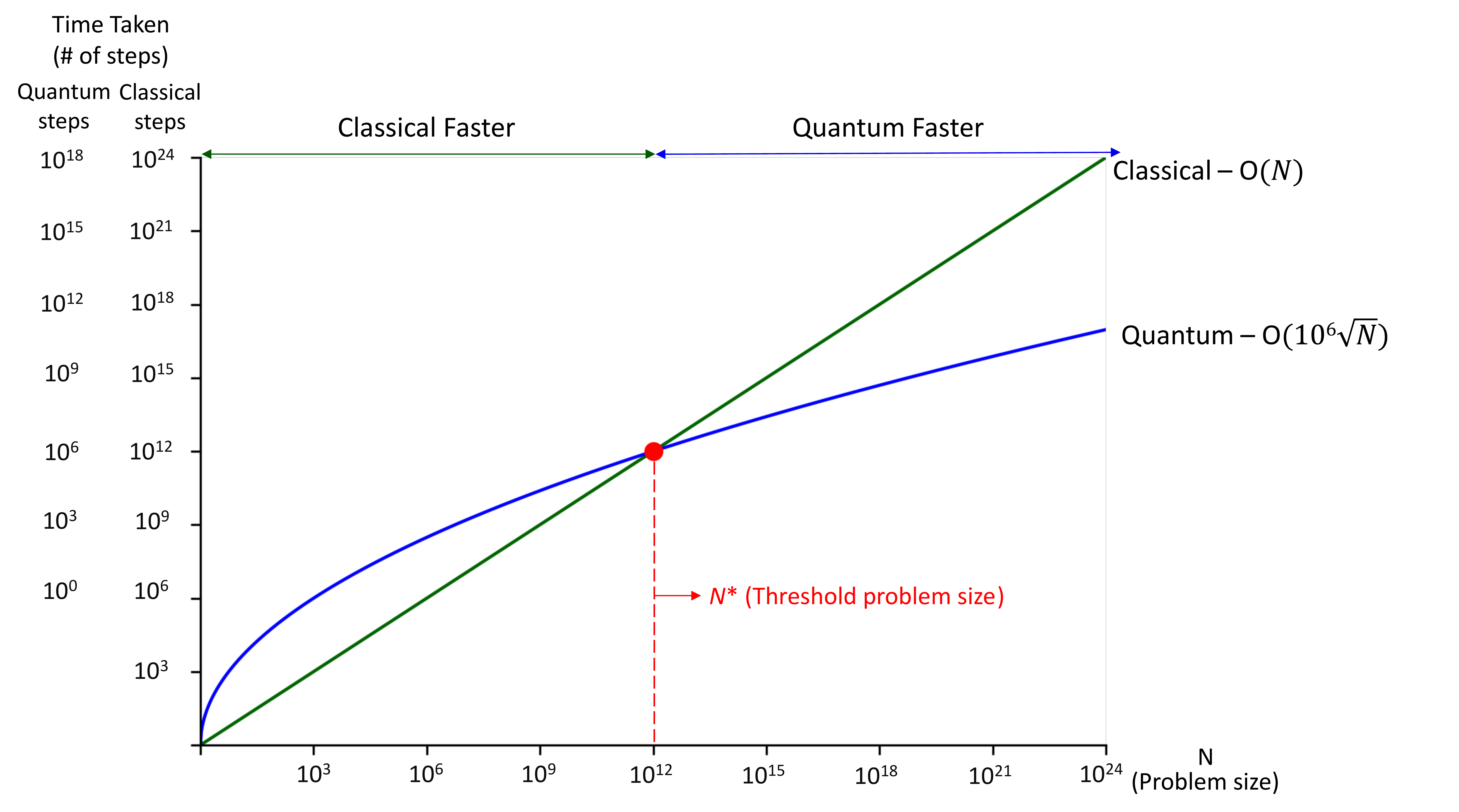}
        \caption{\textbf{Unstructured search:} Application of our framework to the unstructured search problem. The figure depicts the scaling of a quantum algorithm (Grover's) vs. its classical linear counterpart, with a dotted line indicating $N^*$, the threshold problem size where quantum becomes faster than classical. The two notations on the y-axis reflect the speed difference between classical and quantum computers and what they can achieve in a unit of time.}
    \label{fig:example_problem}
\end{figure*}


At the same time, classical computers are much faster. Suppose that in the time that it takes for a quantum computer to do a single operation, a classical computer could do $10^6$ operations (this is derived later in the paper). We thus get an implicit comparison of a classical computer running an algorithm at $N$ and a quantum computer running an algorithm at $10^6\sqrt{N}$. As shown in Figure \ref{fig:example_problem}, this means that for small problem sizes the quantum computer takes much longer. But, as problem sizes get larger, the algorithmic benefit of quantum becomes more important and the two lines cross. This threshold problem size (N*) happens at $10^{12}$. That is, using today's quantum technology, quantum computers would be useful for this problem only at problem sizes larger than $10^{12}$. In one sense this is promising, since human genomes are roughly $10^9$ long, and we have billions of them, so there could be gains for searching within genome libraries. Unfortunately, it also means that quantum computers would need to be able to handle strings of length $10^{12}$ before they are useful for this problem. To search an unstructured database with N elements, Grover's algorithm needs $O(log_2{(N)})$ qubits to represent the indices of the elements in the search space.

To achieve faster performance than a classical computer in this case, $O(\log_2{10^{12}}) \approx 40$ logical qubits are required. Taking the error-correction ratio into account and extrapolating from the IBM hardware roadmap suggests that this milestone would be reached in 2026-2027, when their systems have $40{,}000$ physical qubits (see Figure 3). In actual fact, progress might be slower, for example if qubits grow more slowly or error rates grow. But even with these caveats, our framework suggests that quantum could be useful for genome analysis fairly soon, but only for extraordinarily large searches.

Importantly, this analysis has only included problem-specific constraints for when quantum economic advantage would be achieved. There are broader technical challenges that apply to this problem and many others, which would also need to be solved to implement Grover's algorithm in practice. For example, the setup described above presumes that the data is already loaded in a form accessible to the quantum algorithm. If this were not true, the data would need to be converted as part of the algorithm. 
 But reading the DNA strand will itself take $O(N)$ time alone, bringing the runtime of the quantum algorithm up to $O(N)$, negating any theoretical or practical speedup. However, if one is running many different queries on the same DNA strand, this algorithmic speedup could be beneficial. We discuss this complication in more detail in the Algorithm section.

\subsection{Key inputs}
Our analysis suggests that the practical advantages of a quantum computer for particular problems can be understood, at least broadly, with only a few key inputs:
\begin{enumerate}
    \item The trends/roadmaps for the number of qubits in quantum computers
    \item The number of physical qubits needed per logical qubit (reflecting the quantum error correction needed)
    \item The speed-difference between comparably expensive classical and quantum computers
    \item The computational complexity of classical algorithm
    \item The computational complexity of the quantum algorithm
 \end{enumerate}

Of these, items 1-3 are hardware-level characteristics which only needed to be calculated once per manufacturer for each point in time and then can be used for analysis across many algorithms. We do those calculations, below, and summarize the effective hardware speed difference between quantum and classical computers in section \ref{sec:total_diff}.

Items 4 and 5 are algorithm characteristics.  For classical algorithms, many can be easily sourced from public repositories such as algorithm-wiki.org.  Some quantum algorithms can also be sourced from similar repositories, such as \url{quantumalgorithmzoo.org} or particular research papers. Thus, with the hardware characteristics from 1-3, and the algorithm characteristics from 4 and 5, we have all the pieces needed to apply our framework.

Because our framework applies to the algorithms based on their characteristics, this framework can even be applied to yet-to-be-discovered quantum algorithms because we can calculate how good a quantum algorithm would need to be to outperform current classical alternatives.  That is, we can consider hypothetical examples for item 2 to see how good a quantum algorithm would need to be to good enough for usage. We show the resulting conclusions in Figure 5.

In the sections that follow, we derive each of these items and then discuss the implications for the holistic picture that emerges. In subsequent work, we plan to release a web-based tool that can be used to do these calculations directly.

\section{Hardware Differences}



In this section, we estimate the real-world values of the 3 hardware inputs needed for our framework.

\subsection{Trends in the number of qubits in quantum computers}

\begin{figure}[!ht]
    \centering
    \includegraphics[width=0.47\textwidth, scale=1]{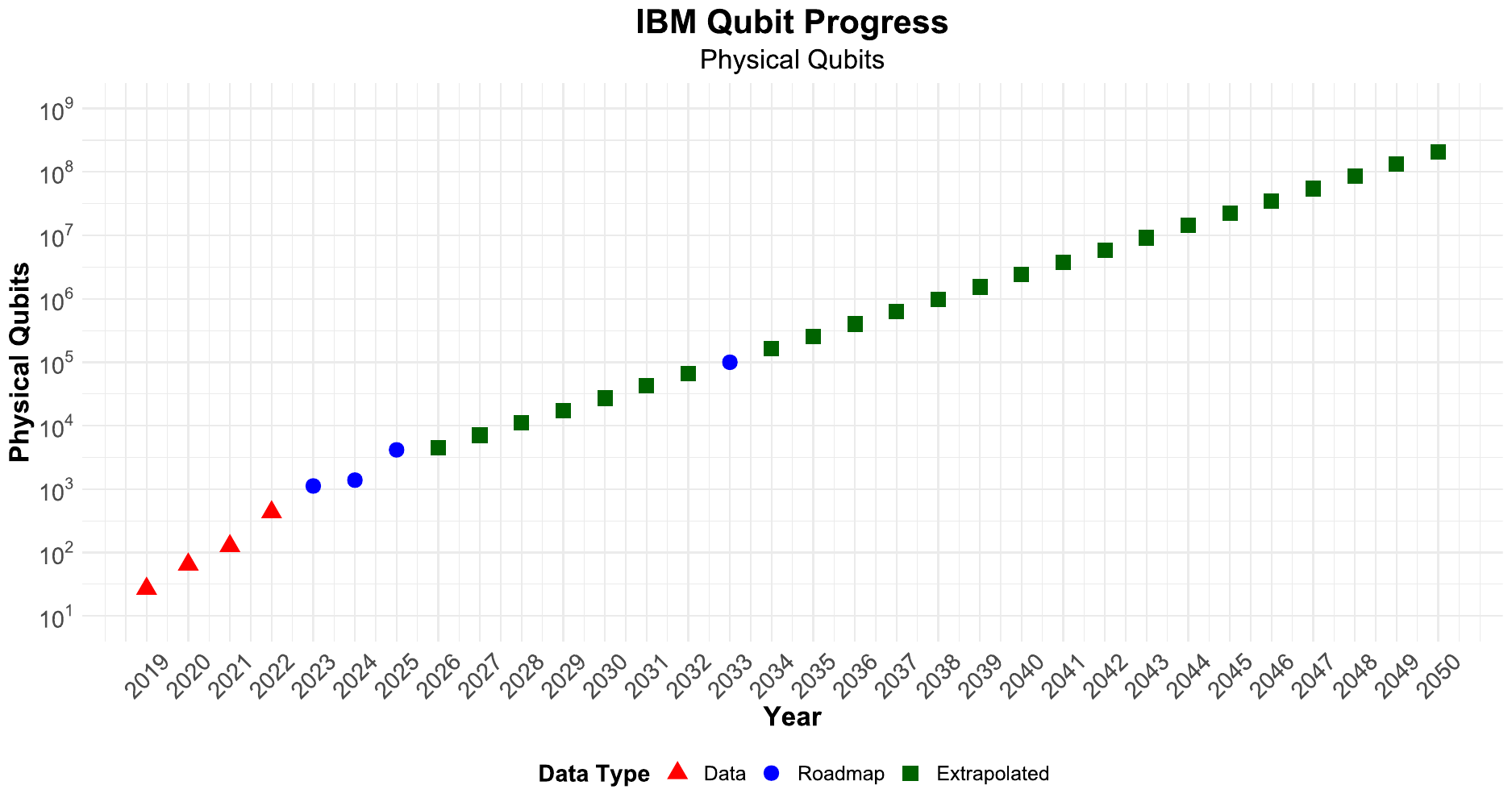}
        \caption{Plot of the number of physical qubits over time. Each color corresponds to the real data (Red), roadmap (Blue), and extrapolated values (Green) based on an exponential growth model.}
    \label{fig:qubit_number_trend}
\end{figure}

To estimate when quantum computers will be sufficiently powerful to solve particular problems, we collect data on planned and realized quantum computers. For the numeric results that follow, we report numbers for IBM's superconducting quantum computers although this same approach could be repeated with other systems, like Google, Rigetti, or others. For example the roadmap of IonQ is shown in the appendix. By 2021, IBM had 127 physical qubits and had roadmapped planned machines with 433 qubits by 2022 and 1,121 qubits by 2023.  Since even these planned systems would not be sufficiently powerful for most applications, so we also extrapolate the recent exponential growth of these systems to higher numbers of qubits (although it is also possible that such systems would face other challenges, like qubit connectivity, explored in \cite{allcock2022does}.). The trend in IBM systems can be seen in Figure \ref{fig:qubit_number_trend}.

Assuming continued exponential growth, IBM will achieve approximately $10^5$ physical qubits by 2033, and $10^{8}$ physical qubits by 2049. Of course, such rapid exponential growth will be very technically challenging and unforeseen problems could slow progress.

\subsection{Error correction: the number of physical qubits per logical qubit}

For most calculations, physical qubits cannot be used as-is, because quantum computers accumulate errors too quickly and thus results calculated this way would often be wrong\cite{lidar2013quantum}. To account for this, quantum error correction can be used, for example those developed by Peres~\cite{peres1985reversible}, Shor~\cite{shor1995scheme}, or Steane~\cite{steane1996multiple}. This requires substantial overhead - recent estimates suggest that 1,000 and 10,000 physical qubits are needed for each logical qubit\cite{Fowler2012, Campbell2017}, although theoretical work also suggests that better alternatives might also be possible \cite{Wood2023}. Error correction also imposes a cost on the number of steps needed for a calculation - a point we revisit later.

The high overhead of physical qubits needed per logical qubit implies that getting sufficient qubits for useful for calculations will take longer.  In particular it implies that IBM will achieve $10^3$ logical qubits by 2038 and $10^6$ logical qubits over 2050.

Because the magnitude of this error correction overhead is so important to when problems become quantum feasible, it is important to consider whether that overhead is changing as (i) systems become larger, or (ii) over time.  We analyze both of these trends in the appendix. We see no systematic changes over time, so we use this $1{:}1000$ overhead ratio for our calculations \cite{Scannell2019}.\footnote{It is plausible that error correction overhead will not increase as quantum computers get more qubits (and hence one can run a larger-sized problem or quantum circuit). This is because error correction can be done to prevent the propagation of errors, regardless of the size of the total circuit. This result is called the threshold theorem, which Neilson and Chuang summarize as ``provided the noise in individual quantum gates is below a certain constant threshold it is possible to efficiently perform an arbitrarily large quantum computation''~\cite{nielsen2002quantum}.}
 
\subsection{The speed difference between classical \& quantum computers}

Quantum computers are much slower than classical computers, hence our analogizing them to tortoises. But how much slower are they? Answering this question has several pieces, so we consider them in turn.

First, consider the raw amount of time it takes for a classical computer to do one classical operation or a quantum computer to do one gate operation. IBM's quantum computers have a gate delay of 555ns\cite{ibm_quantum}\footnote{This is the median two qubit gate time from the date accessed. Two-qubit gates like CNOT are required to enable the speedups in quantum algorithms. Not all computations in a circuit require two-qubit operations, however, and thus may be slightly faster.}, meaning that they run at 2 MHz. This contrasts with current top-end CPUs that run at 5 GHz (e.g., AMD’s Zen 2/3, Intel’s Skylake/Cascade Lake) \cite{intel_cpu, amd_cpu}. On this basis, classical computers are $2{,}500\times$ faster than quantum computers, although one could reasonably modify this up or down somewhat based on whether the classical computer is doing floating point or integer operations.

Rather than calculating a purely-serial speed, one could compare parallel operations (e.g. running on the cloud or a cluster). Quantum computers may permit operations to be run on all qubits at once, which is analogous to a classical processor modifying all data in parallel. Hence we also consider the throughput, or total number of operations that can be done on an entire machine.  For quantum, this is the qubit rate times the number of qubits, whereas for classical computers it is the total processing power of the CPU.\footnote{This is an approximation as it presumes only single-qubit gates matter in a computation. One may also consider the maximum number of multi-qubit operations, which may be also bounded by the connectivity of the qubits in the graph.}

Calculating the number of parallel operations is more complicated, because the right size of classical machine to compare to is unclear. We tackle this issue by calculating parallel operations per dollar. 
IBM offers a cloud-style payment plan at \$1.60 per second. The largest machine IBM offers is the 127 qubit Eagle processor (though technically this is not offered in the on-demand payment scheme and requires a separate contract). Running at a speed of 2MHz, this provides a cost efficiency of $10^{8}$ gate ops per dollar. By contrast, an Amazon c6i.metal classical computer on AWS costs \$5.44/hour, and has a 32 core Intel Xeon Platinum 8375C CPU running at 2.9 GHz, benchmarked at 152GFlops. This provides a cost efficiency of $10^{14}$ FLOPS per dollar. Comparing the two we get a ratio of $10^6$ classical FLOPs to quantum gate ops. That is, on a parallel operations basis, classical computers do $10^6$ more operations per dollar than quantum computers. These calculations suggest that classical computers are $2,500\times$ to $10^6\times$ faster than quantum computers -- with true factor depending on the parallelization of the algorithm.  But there are two complicating factors to this analysis.

The first complicating factor is that these calculations seemingly make the strong assumption that gate operations and flops are equivalently powerful. Objectively this would be wrong, but this issue will actually be resolved by how each set of algorithms is parameterized.  Since classical algorithms are parameterized in classical operations, and quantum algorithms by quantum algorithms, these unit differences will cancel out in the speed calculation.\footnote{In particular, $\#steps \cdot time/step = time$ regardless of whether the step is denoted in gate operations or flops.}

The second complicating factor is quantum error correction and fault tolerance. As mentioned earlier, in addition to the qubit overhead that quantum error correction requires, it also extends the number of calculations that need to be done -- especially to prevent errors from propagating. Here, we approximate overheads from error-correction and fault-tolerance to require running $100$ physical gates to emulate a logical quantum gate. In particular, researchers find a gate over head around 500 for their fault tolerance scheme to succeed at the minimum number of concatenation levels~\cite{chamberland2019fault}. This may be potentially improved by future schemes. However, current error rates may also require multiple concatenation levels, therefore increasing the overhead.


Finally, one might adjust this constant over time, to account for improvements in either quantum or classical hardware. One would expect that in the immediate future, improvements in quantum hardware would outpace improvements in classical hardware, as recently classical hardware has been hitting a wall with the decline of Moore's law, whereas quantum hardware is still in its infancy.

For the calculations that follow, we will use an estimate for the total overhead of $10^6$.

\begin{table*}[]
    \centering

    \begin{tabular}{cccc}
    & Classical & Quantum & Ratio\\
    \toprule
     & \multirow{2}{*}{5GHz\footnotemark[10]}
     & \multirow{2}{*}{2MHz\footnotemark[11]} & \multirow{2}{*}{$10^3-10^4\times$}\\
     Speed& 
     & & \\
     (serial cost) & & \multirow{2}{*}{\color{red} + error-correction overhead [$100\times$] } & \multirow{2}{*}{\color{red} $10^5-10^6\times$}\\
     & &  & \\
    \arrayrulecolor{gray} 
    \midrule
    \multirow{2}{*}{Cost} & \multirow{2}{*}{$10^{14}$ FLOPS/\$\footnotemark[12]} 
    & $10^{8}$ gate ops/\$\footnotemark[13] 
    & $10^6\times$\\
     & & {\color{red} + error-correction overhead [$100\times$]} & {\color{red} $10^8\times$}\\
    \bottomrule
    \end{tabular}

    \caption{Calculation of quantum performance gap, both from cost per serial operation  and cost per (parallel) operation.}
    \label{tab:speeddiffence}
\end{table*}
\footnotetext[14]{Venten et al. (2022) \cite{Velten2022}} 
\footnotetext[15]{Calculated as gate operations per second from single and multi-qubit gate timings. On a Google machine single qubit gate takes 20ns, with a two qubit controlled Z gate taking 32ns \cite{google_quantum}. On an IBM machine, gate time averages are 555ns \cite{ibm_quantum}.}
\footnotetext[16]{\url{https://en.wikipedia.org/wiki/FLOPS}}
\footnotetext[17]{\url{https://www.ibm.com/quantum/access-plans}}

\subsection{The speed difference between classical \& quantum computers - formal version}\label{sec:total_diff}

The net speed advantage of classical computer over a quantum computer can be expressed by a constant, $C$, that reflects the impacts of the factors shown below (Equation \ref{eqn:speed_difference}).

\begin{align}
\small
C = C_\text{speed} * {C_\text{gate overhead}} * {C_\text{alg. constant}} 
\label{eqn:speed_difference}
\end{align}

This includes terms, as follows:
\begin{enumerate}
    \item $C_\text{speed}$: the ratio of the speed of a classical computer divided by the speed of the quantum computer.  As discussed earlier, we assume this is $\approx10,000$.
    \item $C_\text{gate overhead}$: the gate overhead (i.e. additional calculations) that a quantum computer needs to take to maintain its error correction.  We assume this is $\approx 10^2$. 
    \item $C_\text{alg. constant}$: the ratio of the multiplicative constant from the classical algorithm's time complexity divided by that from the quantum algorithm's.\footnote[18]{For example, each classical and quantum linear algorithm would have different constant value for its runtime such as $K_{c}N$ (classical) and $K_{q}\sqrt{N}$ (quantum). On a classical machine this can simply because data structures require more complex operations and therefore cycles to complete. For example an integer divide operation will take far more cycles than an integer add. Similar overheads also exist for quantum computers where sub-computations (represented as circuits) may each require multiple gates. Moreover even the gates in these circuits may not be natively available on a machine and require decomposition into hardware-specific gate sets. One example of this that comes up in practice (largely in Grover's algorithm) is an oracle subroutine, where the typical runtime is described calls to the oracle.
    } Without prior knowledge, the right assumption here is unclear, so we assume this is $1$.
\end{enumerate}

While not included here, one could also imagine additional terms to this equation, reflecting qubit connectivity or slowdowns in running larger problems.


\section{Algorithmic Differences}
Quantum algorithm designers have found a variety of problems in which quantum computers can find solutions asymptotically faster than state-of-the art classical computers. The most famous such algorithm may be Shor's algorithm~\cite{shor1999polynomial} for factoring integers in polynomial time. The difficulty of solving the factoring problem on classical computers has in fact historically been leveraged by cryptography schemes like RSA~\cite{rivest1978method}.\footnote[19]{Modern cryptographic methods have moved away from factoring as a base problem.} One of the first examples of an exponential quantum speed up was found by Deutsch and Jozsa~\cite{Deutsch1992} in 1992, in which the problem was to determine whether a function with a binary output either was constant across all inputs, or had an equal number of true and false results. Other examples of quantum speedups include unstructured search~\cite{grover1996fast}, and a discrete Fourier transform~\cite{coppersmith2002approximate}.

While work in quantum algorithms shows definite promise, not all problems presently benefit from quantum computers equally. This is due to two reasons. First, classical algorithm designers have a substantial head start -- for example Euclid's algorithm for calculating the greatest common denominator is known to have existed since at least 300BC~\cite{brown1971euclid}. Secondly, even in spite of the greater computational model, certain problems may not permit significantly more efficient quantum solutions. As an example, it is not presently known whether quantum computers can solve NP-hard problems, and experts like Scott Aaronson claim ``it seems extremely unlikely that quantum computers can solve NP-complete problems in polynomial time''\cite{aaronson2010bqp}. 

To ensure our analysis of the practical benefit of quantum computing remains generally applicable even with advances in either quantum or classical algorithms, we proprose framing the problem to look at a problem's classical runtime, and investigate what quantum algorithmic runtime would be necessary to overcome the overhead of running on a quantum computer.

As discussed in the section on hardware differences, we will consider an overhead $C=10^6$ classical operations being able to be applied in the same time as one quantum operation. Variants of this analysis with a different overhead are available in the appendix. Given the current classical runtime of an algorithm $f(n)$ and a quantum runtime for an algorithm $g(n)$, we can solve for the minimal problem size $n^*$ at which quantum computing can provide a speedup in the equation $f(n^*) = C g(n^*)$. Note that this doesn't define the amount of the speedup, but rather the minimum size necessary for any speedup to occur. For example, given a problem which has a classical runtime of $O(n^2)$ and a quantum runtime of $O(n)$, the minimum problem size for quantum to provide a speedup would be $n^* = C$. Table~\ref{tab:quoverhead6} plots the value of $n^*$ for several common algorithmic runtimes.

\definecolor{darkgreen}{rgb}{0.6, 1.0, 0.65}
\definecolor{gold}{rgb}{1.0, 0.95, 0.5}
\definecolor{lred}{rgb}{1.0, 0.6, 0.6}
\begin{figure*}[] \label{fig:algorithm_chart}
\large
    \centering
    \begin{subfigure}[b]{\linewidth}
    \centering
    \begin{tabular}{>{\centering\arraybackslash}m{8.5em}|>{\centering\arraybackslash}m{4em}|>{\centering\arraybackslash}m{4em}|>{\centering\arraybackslash}m{4em}|>{\centering\arraybackslash}m{4em}|>{\centering\arraybackslash}m{4em}|>{\centering\arraybackslash}m{4em}}
    \multicolumn{7}{c}{BASE CASE: Classical operations per effective quantum operation=$10^6$}\vspace*{1mm}\\
    \rule{0pt}{1em} \diagbox{Classical}{Quantum} & $\exp n $ & $n^3$ & $n^2$ & $n \log n$ & $n$ & $\log n$\\
     \specialrule{.1em}{0em}{0em} \rule{0pt}{2em}
     \begin{minipage}{4em}\vspace*{-1.5em}\begin{center}$\exp n$\end{center}\end{minipage}\vspace*{-0.5em} & {\cellcolor{lred} $\infty$} & {\cellcolor{darkgreen} 23 } & {\cellcolor{darkgreen} 20 }& {\cellcolor{darkgreen} 18 } & {\cellcolor{darkgreen} 17 }& {\cellcolor{darkgreen} 15} \\
    \hline
    \rule{0pt}{2em} \begin{minipage}{4em}\vspace*{-2em}\begin{center}$n^3$\end{center}\end{minipage}\vspace*{-0.5em}& {\cellcolor{lred} $\infty$}& {\cellcolor{lred} $\infty$} & \cellcolor{gold} $10^6$ & {\cellcolor{darkgreen} $2819$ } & {\cellcolor{darkgreen} $1000$} & {\cellcolor{darkgreen} $173$} \\
    \hline
    \rule{0pt}{2em} \begin{minipage}{4em}\vspace*{-2em}\begin{center}$n^2$\end{center}\end{minipage}\vspace*{-0.5em} & {\cellcolor{lred} $\infty$} & {\cellcolor{lred} $\infty$} & {\cellcolor{lred} $\infty$} & {\cellcolor{gold} $10^7$}  &{\cellcolor{gold}  $10^6$ } & {\cellcolor{darkgreen} $2819$ }\\
    \hline
    \rule{0pt}{2em} \begin{minipage}{4em}\vspace*{-1.75em}\begin{center}$n \log n$\end{center}\end{minipage}\vspace*{-0.5em} & {\cellcolor{lred} $\infty$} & {\cellcolor{lred} $\infty$} & {\cellcolor{lred} $\infty$} &  {\cellcolor{lred} $\infty$} &  {\cellcolor{lred} $10^{434249}$ } & {\cellcolor{gold} $10^6$ }\\
    \hline
    \rule{0pt}{2em} \begin{minipage}{4em}\vspace*{-1.75em}\begin{center}$n$\end{center}\end{minipage}\vspace*{-0.5em} & {\cellcolor{lred} $\infty$} & {\cellcolor{lred} $\infty$}& {\cellcolor{lred} $\infty$} &  {\cellcolor{lred} $\infty$}  & {\cellcolor{lred} $\infty$} & {\cellcolor{gold}  $10^7$ }\\
    \hline
    \rule{0pt}{2em} \begin{minipage}{4em}\vspace*{-1.75em}\begin{center}$\log n$\end{center}\end{minipage}\vspace*{-0.5em} & {\cellcolor{lred} $\infty$} & {\cellcolor{lred} $\infty$} & {\cellcolor{lred} $\infty$} &  {\cellcolor{lred} $\infty$} & {\cellcolor{lred} $\infty$} & {\cellcolor{lred} $\infty$}\\
    \end{tabular}
    \end{subfigure}\\
    \vspace*{1em}
    
    \begin{subfigure}[b]{\linewidth}
    \centering
    \begin{tabular}{>{\centering\arraybackslash}m{8.5em}|>{\centering\arraybackslash}m{4em}|>{\centering\arraybackslash}m{4em}|>{\centering\arraybackslash}m{4em}|>{\centering\arraybackslash}m{4em}|>{\centering\arraybackslash}m{4em}|>{\centering\arraybackslash}m{4em}}
    \multicolumn{7}{c}{OPTIMISTIC CASE: Classical operations per effective quantum operation=$10^4$}\vspace*{1mm}\\
    \rule{0pt}{1em} \diagbox{Classical}{Quantum} & $\exp n $ & $n^3$ & $n^2$ & $n \log n$ & $n$ & $\log n$\\
     \specialrule{.1em}{0em}{0em} \rule{0pt}{2em}
     \begin{minipage}{4em}\vspace*{-1.5em}\begin{center}$\exp n$\end{center}\end{minipage}\vspace*{-0.5em} & {\cellcolor{lred} $\infty$} & {\cellcolor{darkgreen} 18 } & {\cellcolor{darkgreen} 15 }& {\cellcolor{darkgreen} 13 } & {\cellcolor{darkgreen} 11 }& {\cellcolor{darkgreen} 10} \\
    \hline
    \rule{0pt}{2em} \begin{minipage}{4em}\vspace*{-2em}\begin{center}$n^3$\end{center}\end{minipage}\vspace*{-0.5em}& {\cellcolor{lred} $\infty$}& {\cellcolor{lred} $\infty$} & \cellcolor{darkgreen} $10^4$ & {\cellcolor{darkgreen} $234$ } & {\cellcolor{darkgreen} $100$} & {\cellcolor{darkgreen} $33$} \\
    \hline
    \rule{0pt}{2em} \begin{minipage}{4em}\vspace*{-2em}\begin{center}$n^2$\end{center}\end{minipage}\vspace*{-0.5em} & {\cellcolor{lred} $\infty$} & {\cellcolor{lred} $\infty$} & {\cellcolor{lred} $\infty$} & {\cellcolor{gold} $10^6$}  &{\cellcolor{darkgreen}  $10^4$ } & {\cellcolor{darkgreen} $234$ }\\
    \hline
    \rule{0pt}{2em} \begin{minipage}{4em}\vspace*{-1.75em}\begin{center}$n \log n$\end{center}\end{minipage}\vspace*{-0.5em} & {\cellcolor{lred} $\infty$} & {\cellcolor{lred} $\infty$} & {\cellcolor{lred} $\infty$} &  {\cellcolor{lred} $\infty$} &  {\cellcolor{lred} $10^{4342}$ } & {\cellcolor{darkgreen} $10^4$ }\\
    \hline
    \rule{0pt}{2em} \begin{minipage}{4em}\vspace*{-1.75em}\begin{center}$n$\end{center}\end{minipage}\vspace*{-0.5em} & {\cellcolor{lred} $\infty$} & {\cellcolor{lred} $\infty$}& {\cellcolor{lred} $\infty$} &  {\cellcolor{lred} $\infty$}  & {\cellcolor{lred} $\infty$} & {\cellcolor{gold}  $10^6$ }\\
    \hline
    \rule{0pt}{2em} \begin{minipage}{4em}\vspace*{-1.75em}\begin{center}$\log n$\end{center}\end{minipage}\vspace*{-0.5em} & {\cellcolor{lred} $\infty$} & {\cellcolor{lred} $\infty$} & {\cellcolor{lred} $\infty$} &  {\cellcolor{lred} $\infty$} & {\cellcolor{lred} $\infty$} & {\cellcolor{lred} $\infty$}\\
    \end{tabular}    
    \end{subfigure}\\
    
    \vspace*{1em}
    
    \begin{subfigure}[b]{\linewidth}
        \centering
            \begin{tabular}{>{\centering\arraybackslash}m{8.5em}|>{\centering\arraybackslash}m{4em}|>{\centering\arraybackslash}m{4em}|>{\centering\arraybackslash}m{4em}|>{\centering\arraybackslash}m{4em}|>{\centering\arraybackslash}m{4em}|>{\centering\arraybackslash}m{4em}}
    \multicolumn{7}{c}{PESIMISTIC CASE: Classical operations per effective quantum operation=$10^8$}\vspace*{1mm}\\
    \rule{0pt}{1em}  \diagbox{Classical}{Quantum} & $\exp n $ & $n^3$ & $n^2$ & $n \log n$ & $n$ & $\log n$ \\
     \specialrule{.1em}{0em}{0em} \rule{0pt}{2em}
     \begin{minipage}{4em}\vspace*{-1.5em}\begin{center}$\exp n$\end{center}\end{minipage}\vspace*{-0.5em} & {\cellcolor{lred} $\infty$} & {\cellcolor{darkgreen} 28 } & {\cellcolor{darkgreen} 25 }& {\cellcolor{darkgreen} 23 } & {\cellcolor{darkgreen} 21 }& {\cellcolor{darkgreen} 20} \\
    \hline
    \rule{0pt}{2em} \begin{minipage}{4em}\vspace*{-2em}\begin{center}$n^3$\end{center}\end{minipage}\vspace*{-0.5em}& {\cellcolor{lred} $\infty$}& {\cellcolor{lred} $\infty$} & \cellcolor{gold} $10^8$ & {\cellcolor{gold} $10^5$ } & {\cellcolor{darkgreen} $10^4$} & {\cellcolor{darkgreen} $878$} \\
    \hline
    \rule{0pt}{2em} \begin{minipage}{4em}\vspace*{-2em}\begin{center}$n^2$\end{center}\end{minipage}\vspace*{-0.5em} & {\cellcolor{lred} $\infty$} & {\cellcolor{lred} $\infty$} & {\cellcolor{lred} $\infty$} & {\cellcolor{gold} $10^9$}  &{\cellcolor{gold}  $10^8$ } & {\cellcolor{gold} $10^5$ }\\
    \hline
    \rule{0pt}{2em} \begin{minipage}{4em}\vspace*{-1.75em}\begin{center}$n \log n$\end{center}\end{minipage}\vspace*{-0.5em} & {\cellcolor{lred} $\infty$} & {\cellcolor{lred} $\infty$} & {\cellcolor{lred} $\infty$} &  {\cellcolor{lred} $\infty$} &  {\cellcolor{lred} $10^{43429448}$ } & {\cellcolor{gold} $10^8$ }\\
    \hline
    \rule{0pt}{2em} \begin{minipage}{4em}\vspace*{-1.75em}\begin{center}$n$\end{center}\end{minipage}\vspace*{-0.5em} & {\cellcolor{lred} $\infty$} & {\cellcolor{lred} $\infty$}& {\cellcolor{lred} $\infty$} &  {\cellcolor{lred} $\infty$}  & {\cellcolor{lred} $\infty$} & {\cellcolor{gold}  $10^9$ }\\
    \hline
    \rule{0pt}{2em} \begin{minipage}{4em}\vspace*{-1.75em}\begin{center}$\log n$\end{center}\end{minipage}\vspace*{-0.5em} & {\cellcolor{lred} $\infty$} & {\cellcolor{lred} $\infty$} & {\cellcolor{lred} $\infty$} &  {\cellcolor{lred} $\infty$} & {\cellcolor{lred} $\infty$} & {\cellcolor{lred} $\infty$}\\
    \end{tabular}
    \end{subfigure}
    \vspace*{1em}
    \caption{The minimum problem size necessary for a problem size given standard (a), optimistic (b), and pessimistic (c) assumptions of an overhead of $10^6$, $10^4$, and $10^8$, respectively. An additional calculation using $10^3$ is shown in the appendix. Cells with a red fill denote that the classical computer will always win in practice. Cells with a yellow fill denote algorithms which require a problem size of more than $10^5$ for a quantum computer to be faster. Cells with a yellow fill denote algorithms which require a problem size less than $10^5$ for quantum computers to achieve a speedup. Note that here we model $\exp(n)$ as $2^{n}$. }
    \label{tab:quoverhead6}
\end{figure*}

Due to the overhead of quantum hardware, in order to achieve a speedup, the quantum algorithm for solving a problem must be strictly more efficient than the best classical algorithm for solving the problem for $n^*$ to be finite\footnote[20]{Since the qubit on a quantum computer can be used to emulate a classical bit, the best 
 quantum algorithm for any problem is at least as fast (but perhaps equivalent in runtime) as a classical algorithm for the problem.}. Problems with exponentially faster quantum algorithms have small minimum problem sizes if the classical algorithm takes greater than $O(n^2)$ time. For problems with a linear or greater reduction in runtime, have a threshold problem of at most $C$. Any problems, with a sublinear speedup, however, have a threshold problem size of greater than $C$ -- often much greater.

Let us now consider this from the point of view of a practitioner with a particular problem in mind. If your current algorithm takes exponential time to run, this may be a good candidate for a quantum speedup -- assuming a polynomial or faster exponential quantum algorithm can be found.\footnote[21]{For example, reducing classical runtime from $\exp 2n$ to a quantum runtime of $\exp n$.} This becomes more ambiguous, however, if your classical algorithm has a polynomial runtime. Depending the value of $C$, you likely will need a speedup of at least $n$ on a quantum computer for reasonable threshold problem sizes. If your algorithm's runtime has a high order (e.g. $n^4$ or $n^3$), finding such an algorithm may be plausible. In contrast, if the classical algorithm you use has a lower order (e.g. $n \log n$), reducing the runtime by $n$ to $\log n$ is equivalent to finding an exponentially faster quantum algorithm. The runtime of many common classical problems with $n$ taken as the conventional input size has been summarized in Appendix Table~\ref{tab:classprob}.

We can also consider this analysis from the perspective of finding a quantum algorithm with a speedup. Shor's factoring algorithm runs in polynomial time, whereas the best classical algorithm in nearly exponential time. From our analysis, this shouldn't require a very large problem size to merit a speedup. 
%
In contrast, the speedup of unstructured search from $O(N)$ to $O(\sqrt N)$ with Grover's algorithm, requires a minimum problem size of $n^*=C^2$, which using $C=10^6$  means $n^*=10^{12}$. The Quantum Fourier Transform, with a runtime of $O(\log^2 n)$ -- an exponential speedup over the classical $O(n \log n)$ Fast Fourier Transform -- requires $n^*=10^7$ to merit a speedup. The HHL algorithm~\cite{Harrow_2009} provides a speedup on the runtime of solving a sparse linear system of equations (assuming constant condition number and error tolerance) from $O(n)$ to $O(\log n)$, requiring a size of $n^*=10^7$.

\section{Sensitivity testing}
To get the answers in the previous section, we estimated several parameters (speed difference, error correction overhead, etc.) and evaluation metrics (cost per serial operation and cost per parallel operation). In this section we consider how much our conclusions change if we assume more optimistic (or pessimistic) estimates.

\subsection{Optimistic version}
Here we consider how the thresholds for quantum economic advantage change if we make more optimistic assumptions for the performance of quantum computers, resulting in performance gap of $10^4$. For example, this might arise if:
\begin{enumerate}
    \item error correction codes improve,
    \item quantum hardware produces fewer errors,
    \item the costs of building and operating quantum computers falls more rapidly than classical computers
\end{enumerate}

This optimistic version is shown at the middle of \autoref{tab:quoverhead6}. The overall results are qualitatively similar to the base case, but certain problems (e.g. when classical is $n^3$ and quantum is $n$), become attractive for problem sizes of only a few hundreds, that could be quite business-relevant.

\subsection{Pessimistic version}
Here we consider what thresholds for quantum economic advantage that we would predict if we instead made more pessimistic assumptions\footnote[22]{This does not consider improvements in algorithms, that would lead to problems changing complexity classes, such as if quantum computers could be efficiently simulated.} that lead to a quantum performance gap of $10^8$. This might arise if:
\begin{enumerate}
    \item error correction becomes harder as systems get larger,
    \item classical computers improve their price-performance faster than classical computers,
    \item the connectivity issues for quantum computers make each logical qubit require more effective classical qubits
\end{enumerate}

This pessimistic version is shown at the bottom of \autoref{tab:quoverhead6}. The overall results are again qualitatively similar to the base case, but now require problem sizes in the hundreds of millions or billions, putting many everyday business applications out of reach. 

\section{Limitations to the analysis}

\subsection{Contextualizing Problem Size}
It is important to note that these problem sizes $n$ vary on the context of the problem. As a result, a very large $n$ may be common in some problems, or uncommon in others. For example, we describe the runtime of Shor's algorithm as polynomial in $n$, where $n$ is used to denote the number of bits of the value being factored. Therefore, a value of $n=20$, denotes factoring a number of size $2^{20}\approx10^6$. Thus if the using the analysis above you derived a threshold problem size of $n^*=20$ in bits, that would correspond to $2^{n^*}$ being the threshold problem size in terms of the value of the integer being factored.

Understanding this conversion may often be important to practitioners. As an example, while the $n^*=C^2=10^{12}$ of may appear too large for most use cases to search through a list of size $n$, Grover's algorithm may be used to search through an exponentially large ``list'' of possible solutions to a problem. For example, one could apply Grover's algorithm to perform an exhaustive search over all $n=2^m$ settings of a satisfiability problem of $m$ variables. Grover's quadratic speedup on such a search problem would then result in $2^{m/2}$ evaluations of the search function~\cite{montanaro2016quantum}. The threshold problem size above would remain correct at $n^*=10^{12}$, that would correspond to a threshold number of variables of $m^*=\log_2 n^* \approx 40$.


\subsection{Precision}
When analyzing the very small minimum problem sizes like in an algorithmic improvement from exponential time to polynomial, our order-of-magnitude analysis of the expected problem size becomes less accurate. This is because the additional constant overheads within the algorithms themselves, including constructing the correct quantum gateset or classical function, becomes more crucial. For example, when analyzing Shor's algorithm under our order-of-magnitude framework with an overhead constant $C=10^6$, we find that factoring numbers of only a few bits will achieve speedup on a quantum computer. To determine more than rough estimates in these regime, significant consideration must be made for the the quantum-gate overhead constants, as well as lower-order constant overheads for the classical algorithms.  

\subsection{Quantum RAM / Data Loading}

Thus far in our analysis, we have implicitly assumed that computing time will be the bottleneck to any computational task and thus ignored the time for loading the data. But this assumption can be wrong if the time to load the data is significantly longer than the time needed for the calculation. In these cases, the bottleneck will be the loading time of the data. This problem can arise particularly for quantum computers because of how they represent data or need to convert classical data into a quantum format.

In classical computing, doing an operation on one bit of information requires loading (at most) one bit from memory. Therefore, the cost of loading data is (at most) proportional to the number of operations. And thus, data loading costs cannot change the asymptotic scaling of classical algorithms \footnote[23]{Assuming efficient ``random-access'' implementations that don't thrash, etc.}. By contrast, in quantum computing each operation acts on a qubit, whose superposition might require exponentially many bits of classical information to construct. In this case, the loading operations needed to build up the superpositions could be much greater than the operations needed for a calculation. In particular, if the amount of information needed to construct the superposition grows as the problem size grows, then the scaling of loading data could be worse than that of the calculation itself.



``Quantum RAM (qRAM)''~\cite{giovannetti2008quantum} aims to provide a way of storing quantum information for later use that does not require conversion between classical and quantum formats, and hence avoids making the load operations too expensive.

Thus, loading data on a quantum computer can itself be the limiting factor on the scaling of a quantum algorithm if the data usage computations scale less well than the calculations on that data. It can also be a bottleneck if the data is stored in classical format, in which case there may be a computational penalty needed to convert it into a quantum-machine readable format. In this case, the time complexity of loading or storing data could be significantly worse than linear.



In practice, the time needed for loading thus becomes a lower bound on how well the quantum algorithm can do. For example, as we discussed with the DNA database example, an algorithm with a time complexity of $O(\sqrt{n})$ but where data loading is linear $O(n)$ will be limited by data loading. As such, an apparent computational benefit that a quantum algorithm has, might be erased by this effect.

\subsection{Qubit connectivity}

In our analysis so far, we have assumed that, once the error-correction overhead is paid, physical qubits can perfectly represent logical qubits. This assumption is optimistic because real qubits have limited connectivity, unlike their fully-connected theoretical counterparts. Sometimes this distinction can be hidden by clever quantum compilers that select and re-route computations such that the necessary qubits are adjacent~\cite{10.1145/3520434,cowtan_et_al:LIPIcs:2019:10397}. Solving for the optimal qubit assignment has been shown to be NP-complete~\cite{10.1145/3168822}, with approximation algorithms, like dynamic programming, being possible practical solutions. Such approximations can result in orders-of-magnitude additional costs~\cite{10.1145/3168822}. 


\section{Discussion}
Our findings show that quantum computers will outpace classical computers when (i) there are significantly better quantum algorithms and (ii) the size of the problem being solved is sufficiently large. This latter finding is particularly important because near-term quantum computers have limited computational capacity and thus the need to run on a large problem is synonymous needing to wait for longer for quantum computers to get powerful enough to tackle these problems.

Overall, we find two rules useful to practitioners for determining whether quantum can provide a speedup. If a classical algorithm takes exponential time and there exists a polynomial quantum algorithm, you're likely to get a speedup. Additionally, if the quantum algorithm is significantly better, say by a factor of $n$, and the problem size being tackled is larger than speed difference between classical and quantum computers --- for example if classical computers are $1{,}000{,}000\times$ faster then quantum computers will be better if the problem being tackled must have $n > 1{,}000{,}000$.

While there are several examples of quantum algorithms that provide speedups beyond the best-known classical algorithms, it is currently unclear what the precise advantages and limitations are at an algorithmic level. As of this text, it has not been proven that the set of problems solvable by quantum computers in polynomial time (more specifically, the BQP complexity class) contains any problem which is not in the corresponding complexity class for classical computers (BPP). More concretely, while there are some problems like factoring which are known to have a polynomial time quantum algorithm, it has not been proven that a polynomial time algorithm for factoring cannot exist for classical computers.

Future study can compare and contrast classical and quantum algorithms to determine which ones need to be developed. Furthermore, while our analysis relies on established quantum error correction rates, it is important to note that our results could vary depending on the progress and advancements in the field of quantum error correction~\cite{Wood2023}. Future study can improve this. 

Lastly, this study can be extended to industry applications. For example, several industries such as Automotive, Chemistry, and Finance have started to examine an application of quantum computing \cite{Egger2020, Elfving2020, Luckow2021}. Our analysis provides a framework for practitioners such as these to analyze actual quantum algorithms and when they'll be useful, as well as prospective quantum algorithms analyzed as part of scenario planning.


%


\newpage

\section*{Appendix}

\renewcommand\thefigure{\arabic{figure}}    
\setcounter{figure}{0}

\begin{figure}[H]
    \centering
    \includegraphics[width=0.47\textwidth, scale=1]{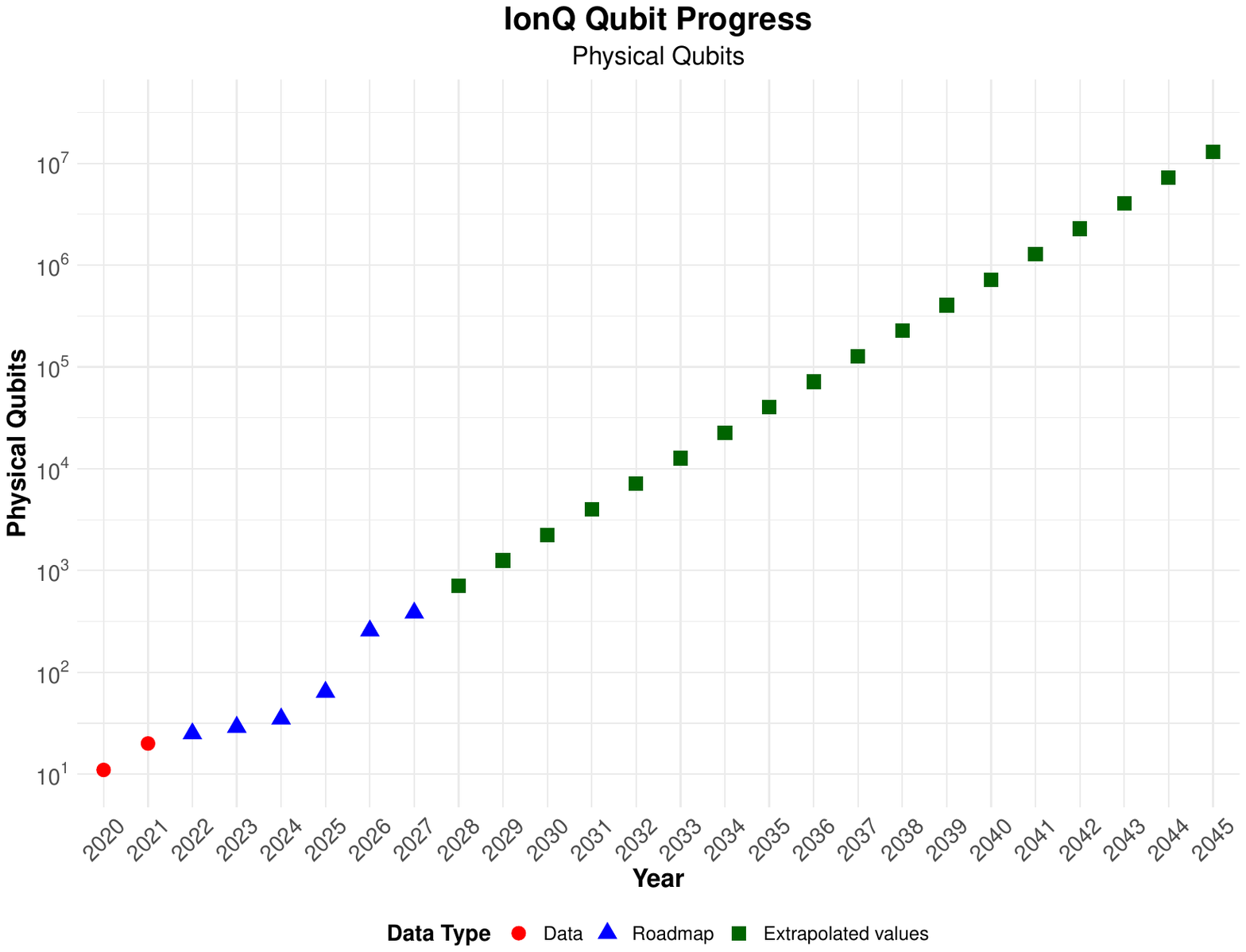}
        \caption{Plot of the number of physical qubits over time. Each color corresponds to the current data (Red), roadmap (Blue), and extrapolated values (Green) based on an exponential growth model.}
    \label{fig:IonQ2}
\end{figure}

Appendix Figure 1 displays the trend of the number of qubits in IonQ devices over time. Based on the extrapolation, IonQ is projected to reach approximately $10^3$ physical qubits by 2030 and $10^{6}$ physical qubits by 2040, as illustrated in the figure.

Appendix Figure 2 displays the physical qubit error rate (Pauli X error) over time. To obtain this data, we gathered information on IBM's quantum processors\footnote[24]{https://quantum-computing.ibm.com/services?services=systems}. 

\begin{figure}[H]
    \centering
    \includegraphics[width=0.485\textwidth]{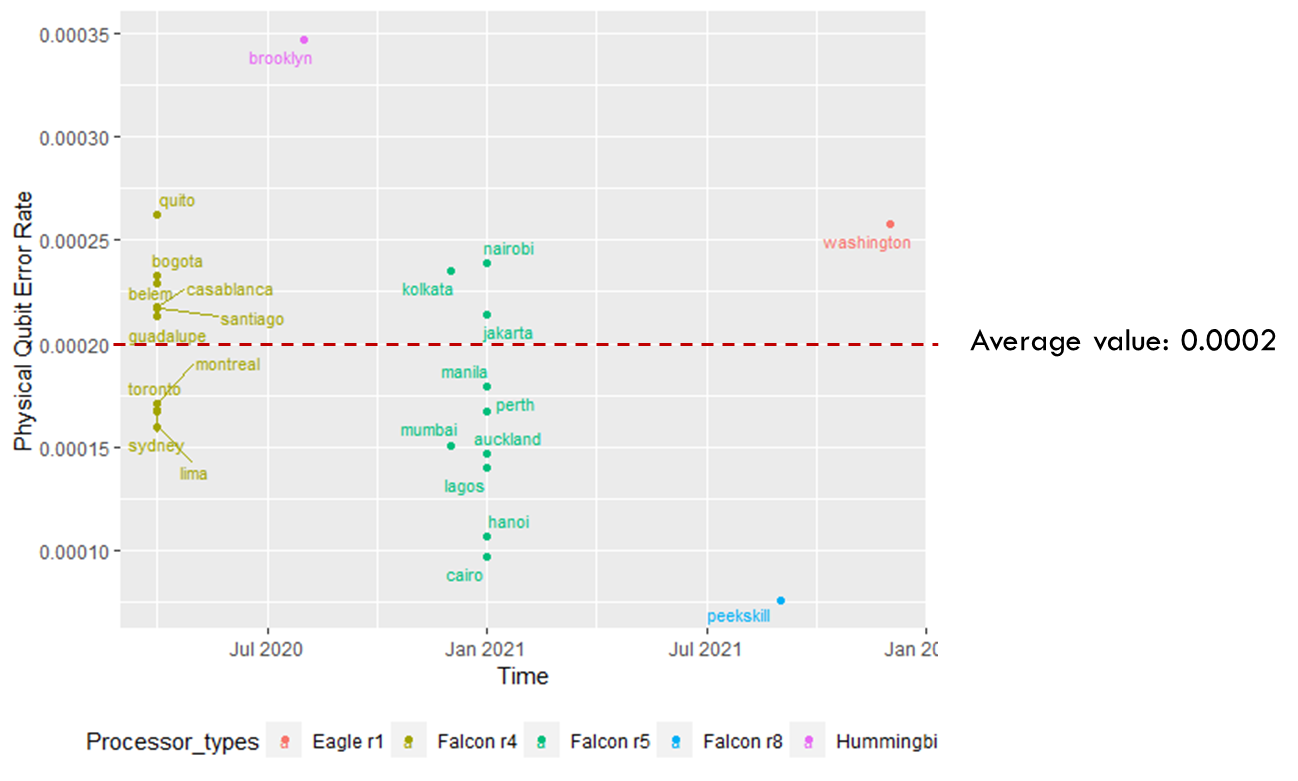}
    \caption{Plot of Error Rates over Time}
    \label{fig:my_label}
\end{figure}

We illustrate the relationship between the number of physical qubits and their error rate to evaluate qubit effectiveness in Appendix Figure 3. To achieve this, we collected data from IBM quantum processors and found that, in some cases, the error rate increases with the number of qubits. However, this correlation is not statistically significant. Instead, we discovered that physical qubit error rates appear to be well-managed, as the error has not grown proportionally with the increase in the number of qubits.

\begin{figure}[H]
    \centering
    \includegraphics[width=0.47\textwidth]{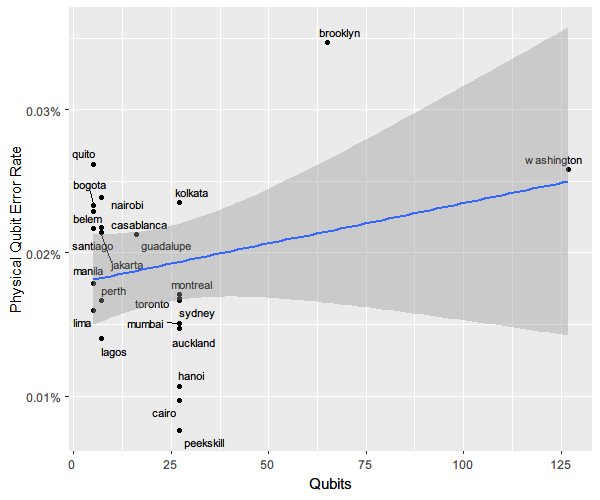}
    \caption{The number of physical qubits and error rate}
    \label{fig:my_label}
\end{figure}

\begin{figure}[h]
    \centering
    \includegraphics[width=0.47\textwidth]{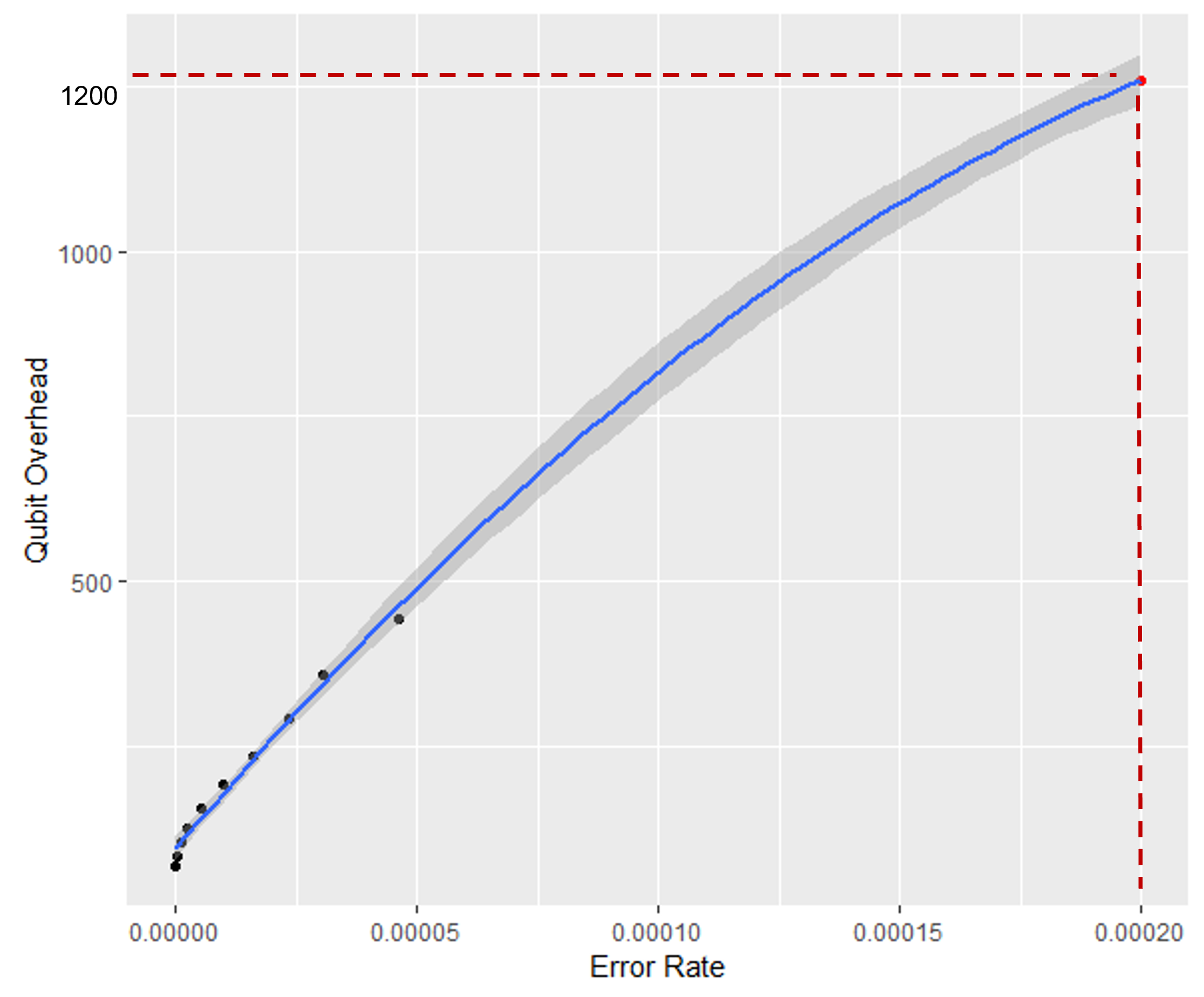}
    \caption{Plot of the qubit overhead and error rate}
\end{figure}

Utilizing a recent simulation study's error rate data \cite{Chamberland2020}, we sought to identify a relationship between the qubit overhead and error rate for physical qubits. Appendix Figure 4 reveals a positive correlation between qubit overhead and error rate. When we incorporate observational data from IBM into this overhead trend, the average error rate is 0.0002 (see Figure 2 in the Appendix), resulting in an overhead greater than 1,200 when using this value. Consequently, the quantum feasibility of current hardware might not be adequate for addressing large-scale problems.

\setcounter{table}{0}   
\begin{table}[H]
    \centering
    \resizebox{\columnwidth}{!}{%
    \begin{tabular}{c|l}
    \multicolumn{2}{c}{Classical Algorithm}\\
    \hline
    \multirow{2}{*}{Exp} & 3-Graph /  4-Graph Coloring (NP) \\
     & Subset-Sum (NP)\\
     & Vertex Cover (NP) \\
     & Set-Covering (NP) \\
     & Maximum Cut (NP) \\
     & The Traveling-Salesman Problem (NP) \\
     & Enumerating Maximal Cliques (NP) \\
     & n-Queens (NP) \\
     & Graph Edit Distance Computation (NP~\cite{zeng2009comparing}) \\
     & Turnpike\\
     & NFA to DFA Conversion  \\
     & Dependency Inference \\
     & BCNF Decomposition  \\
     & 4NF Decomposition  \\
     & Discovering Multivalued Dependencies \\
     & Finding Frequent Item Sets \\
     & Motif Search \\
     & Minimum Wiener Connector \\
     & Change-Making \\
     & Median String \\
     & Factoring\\
    \hline
    \multirow{2}{*}{$n^4$} & Longest Path on Interval Graphs  \\
     & Determinant using Integer Arithmetic \\
    \hline
    \multirow{2}{*}{$n^3$} & Grobner Bases \\
     & Maximum-weight Matching \\
     & MDPs for Optimal Policies  \\
     & Link Analysis (Indegree) \\
     & Integer Relation  \\
    \hline
    \multirow{2}{*}{$n^{2.38}$} & Parsing \\
     &  \\
    \hline
    \multirow{2}{*}{$n^{2.188}$} & Transitive Reduction \\
     &  \\
    \hline
    \multirow{2}{*}{$n^2 \log n$} & Nash Equilibria  \\
     & Entity Resolution \\
     & 2-D Elliptic Partial  \\
     & 3-D Elliptic Partial  \\
     & All-pairs Shortest Path  \\
    \hline
    \multirow{2}{*}{$n^2$} &  Factorization of Polynomials over Finite Fields\\
     & Cryptanalysis of Linear Feedback Shift Registers \\
     & Translating Abstract Syntax Trees into Code \\
     & Digraph Realization \\
     & Sequence to Graph Alignment (Linear Gap Penalty) \\
     & Rod Cutting \\
     & Frequent Words\\
    \hline
    \multirow{2}{*}\tiny{$n^2/\log n$} &  Sequence Alignment  \\
    \hline
    \multirow{2}{*}\tiny{$n^2/\log^2 n$} &  Longest Common Subsequence  \\
    \hline
    \multirow{2}{*}{$n^{1.5}$} & Max Flow \\
     &  \\
    \hline
     \multirow{2}{*}{$n^{1.188}$} & Maximum Cardinality Matching \\
     &  \\
    \hline
    \multirow{2}{*}{$n^{1.186}$} & Matrix Multiplication \\
     &  \\
    \hline
    \multirow{2}{*}{$n^{1.185}$} & Linear Programming \\
     &  \\
    \hline
    \multirow{2}{*}\tiny{$n \log n \log \log $} & Greatest Common Divisor \\
    \hline
    \end{tabular}
    }
    \caption{Classical runtime of various computer science problems, with $n$ taken as the conventional input size for the problem (Continued).}
\end{table}

\setcounter{table}{0} 
\begin{table}[H]
    \centering
    \resizebox{\columnwidth}{!}{%
    \begin{tabular}{c|l}
    \multicolumn{2}{c}{Classical Algorithm}\\
    \hline
    
     \multirow{2}{*}{$n \log n$} & LU Decomposition  \\
     & Multiplication \\
     & Discrete Fourier Transform  (QEXP) \\
     & Cyclopeptide Sequencing  \\
     & Comparison Sorting \\
     & Line Segment Intersection  \\
     & Convex Hull  \\
     & Closest Pair Problem \\
     & SDD Systems Solvers \\
     & Polygon Clipping \\
     & Nearest Neighbor Search \\
     & Voronoi Diagrams \\
     & Delaunay Triangulation \\
     & Weighted Activity Selection  \\
     & Single-interval Scheduling Maximization (Unweighted)\\
    \hline
    \multirow{2}{*}{$n \log \log n$} & Shortest Path \\
     & Duplicate Elimination \\
    \hline
    \multirow{2}{*}{$n$} & Line Clipping  \\
     & Integer Sorting \\
     & kth Order Statistic \\
     & Linear System of Equations \\
     & Strongly Connected Components \\
     & Minimum Spanning Tree  \\
     & String Search \\
     & Joins \\
     & Cycle Detection \\
     & Generating Random Permutations \\
     & Minimum value  \\
     & All Permutations \\
     & Huffman Encoding \\
     & Constructing Eulerian Trails in a Graph \\
     & Line Drawing \\
     & Topological Sorting \\
     & DFA Minimization  \\
     & Lowest Common Ancestor \\
     & De Novo Gene Assembly \\
     & Disk Scheduling \\
     & Lossy Compression\\
     & Stable Marriage \\
     & Maximum Subarray\\
     & Constructing Suffix Trees \\
     & Longest Palindrome Substring\\
     & Matrix Factorization for Collaborative Filtering \\
     & Point in Polygon \\
     & Polynomial Interpolation \\
     & Deadlock Avoidance\\
     & Page Replacements\\
     & Recovery\\
    \hline
    \multirow{2}{*}{$\log n$} & Mutual Exclusion  \\
     & Self-Balancing Trees Insertion  \\
     & Self-Balancing Trees Deletion  \\
     & Self-Balancing Trees Search  \\
    \hline
    \end{tabular}
    }
    \caption{Classical runtime of various computer science problems, with $n$ taken as the conventional input size for the problem.}
    \label{tab:classprob}
\end{table}

\begin{table}[H]
    \centering
    \resizebox{\columnwidth}{!}{%
    \begin{tabular}{c|c|c|c|c|c|c}
    \multicolumn{7}{c}{Quantum Algorithm Runtime}\vspace*{2mm}\\
    Classical & $\exp n $ & $n^3$ & $n^2$ & $n \log n$ & $n$ & $\log n$\\
    \hline
    $\exp n$ & $\infty$ & 15 & 12 & 10  & 9 & 8\\
    $n^3$ & $\infty$ & $\infty$ & $10^3$ & $65$  & $32$ & $14$\\
    $n^2$ & $\infty$ & $\infty$ & $\infty$ & $9118$  & $10^3$ & $65$\\
    $n \log n$ & $\infty$ & $\infty$ & $\infty$ & $\infty$  & $10^{434}$ & $10^3$\\
    $n$ & $\infty$ & $\infty$ & $\infty$ & $\infty$  & $\infty$ & $9118$\\
    $\log n$ & $\infty$ & $\infty$ & $\infty$ & $\infty$  & $\infty$ & $\infty$\\
    \end{tabular}
    }
    \caption{Minimum problem size necessary for a quantum speedup, given each quantum operation has an overhead of $C=10^3$ compared to a classical operation.}
    \label{tab:quoverhead3}
\end{table}

\section*{Acknowledgment}
The authors would like to thank Francesco Bova, Tom Lubinski, Jayson Lynch, Bill Kuszmaul, William D. Oliver, and the MIT SuperTech group for thoughtful discussions on quantum computing and its implications. We also appreciate the thoughtful feedback from seminar audiences at the Economics of Quantum Information Technology, the FutureTech-CSET workshop, the Transatlantic Quantum Forum, and the Quantum Economic Development Consortium (QED-C) Seminar. This project is supported by Accenture. We extend our sincere appreciation to the Accenture team, including individuals such as Prashant P. Shukla, Carl M. Dukatz, Laura G. Converso, and Kung-Chuan (KC) Hsu for their valuable collaboration. Sukwoong Choi and Neil Thompson are supported by funding from Good Ventures and Accenture. William S. Moses was supported in part by a DOE Computational Sciences Graduate Fellowship DE-SC0019323. This research was supported in part by Los Alamos National Laboratories Grant 531711. This material is based upon work supported by the Department of Energy, National Nuclear Security Administration under Award Number DE-NA0003965.

\ifCLASSOPTIONcaptionsoff
  \newpage
\fi



%

\bibliographystyle{IEEEtran}
\bibliography{references}

\newpage

\begin{IEEEbiography}[{\includegraphics[width=1in,height=1.25in,clip,keepaspectratio]{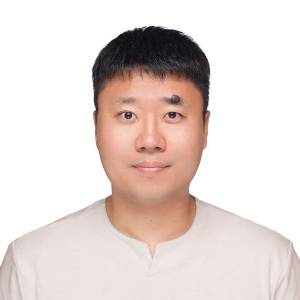}}]{Sukwoong Choi} is an assistant professor at the Department of Information Systems and Business Analytics, School of Business, in the University at Albany, SUNY. He is also a Digital Fellow at Initiative on the Digital Economy (IDE) of MIT. Before joining the University at Albany, he was a postdoctoral scholar at MIT Sloan School of Management and MIT Computer Science and Artificial Intelligence Lab (CSAIL), the University of Southern California (Technology Innovation and Entrepreneurship), and the University of Kentucky (Gatton College of Business and Economics, Management). He has a PhD from the School of Business and Technology Management from KAIST.
\end{IEEEbiography}
\begin{IEEEbiography}
[{\includegraphics[width=1in,height=1.25in,clip,keepaspectratio]{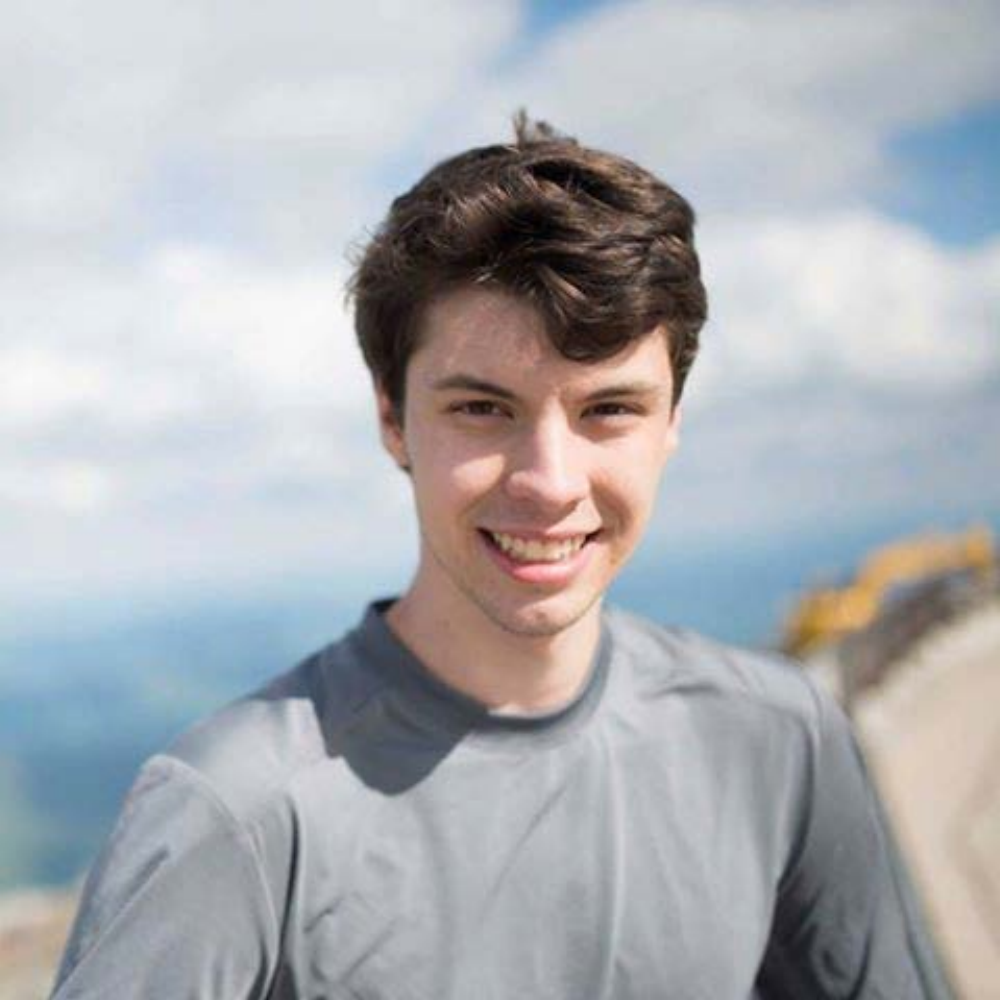}}]{William S. Moses} is an assistant professor at University of Illinois Urbana Champaign in the Computer Science department, starting in 2024. Previously, he was a PhD candidate and undergrad at MIT studying computer science and physics, and a J. Tinsley Oden Faculty Fellow at the University of Texas, Austin. He is a recipient of the U.S. Department of Energy Computational Science Graduate Fellowship and the Karl Taylor Compton Prize, MIT's highest student award.
\end{IEEEbiography}
\begin{IEEEbiography}
[{\includegraphics[width=1in,height=1.25in,clip,keepaspectratio]{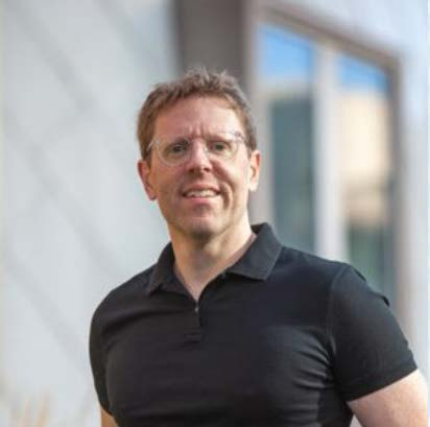}}]{Neil Thompson} is the Director of the FutureTech research project, where his group studies the economic and technical foundations of progress in computing, and is cross-appointed at MIT’s Computer Science and AI Lab and MIT’s Initiative on the Digital Economy. Previously, Neil was an Assistant Professor of Innovation and Strategy at the MIT Sloan School of Management, where he co-directed the Experimental Innovation Lab (X-Lab), and a Visiting Professor at the Laboratory for Innovation Science at Harvard. He has a PhD in Business and Public Policy from Berkeley, where he also did Masters degrees in Computer Science and Statistics. He also has a masters in Economics from the London School of Economics, and undergraduate degrees in Physics and International Development. 
\end{IEEEbiography}

%




\end{document}